\documentclass[useAMS,usenatbib]{mn2e}
\usepackage{graphicx}
\usepackage{grffile}
\usepackage{lscape}
\usepackage{verbatim}
\title[Two new ultra-cool benchmark systems from WISE+2MASS]{Two new ultra-cool benchmark systems from WISE+2MASS}
\author[Gomes et al.]{J. I.~Gomes,$^1$ D. J.~Pinfield,$^1$ F. Marocco,$^1$ A.C.~Day-Jones,$^{1,2}$ B.~Burningham,$^1$ \\ \newauthor Z. H. Zhang,$^1$ H.R.A.Jones,$^1$ L. van Spaandonk,$^1$ D. Weights$^1$ \\ $^1$Centre for Astrophysics Research, University of Hertfordshire, College Lane, Hatfield, Hertfordshire, AL10 9AB.\\$^2$Departamento de Astronomia, Universidad de Chile, Camino del Observatorio 1515, Santiago, Chile.}

\begin{document}

\date{}

\pagerange{\pageref{firstpage}--\pageref{lastpage}} \pubyear{2012}

\maketitle

\label{firstpage}

\begin{abstract}
We have used the 2MASS all-sky survey and the WISE to look for ultracool dwarfs that are part of multiple systems containing main sequence stars. We cross-matched L dwarf candidates from the surveys with Hipparcos and Gliese stars, finding two new systems. The first system, G255-34AB is an L2 dwarf companion to a K8 star, at a distance of 36 pc. We estimate its bolometric luminosity as log $\mathit{L/L_{\odot}}$ = $\mathit{-}$3.78 $\pm$ 0.045 and $T_{\rm eff}$ = 2080 $\pm$ 260 K. The second system, GJ499ABC, is a triple, with an L5 dwarf as a companion to a binary with an M4 and K5 star. These two new systems bring the number of L dwarf plus main sequence star multiple systems to twenty four, which we discuss. We consider the binary fraction for L dwarfs and main sequence stars, and further assess possible unresolved multiplicity within the full companion sample. This analysis shows that some of the L dwarfs in this sample might actually be unresolved binaries themselves, since their $\mathit{M_J}$ appears to be brighter than the expected for their spectral types.  
\end{abstract}


\section{Introduction}
In the last years an increasing number of binary systems with ultracool components have been discovered and studied. Ultracool dwarfs (UCDs) are defined as dwarf stars with spectral types of M7 or later, and include substellar brown dwarfs (\citealt{baraffe1998,jones2001}; masses $<$ 0.072 M$_\odot$) and some very low mass stars. UCDs populate the lower temperature range, from effective temperatures $T_{\rm eff}$ $\sim$ 2400 K to 500 K \citep{kirkpatrick2005}. 

Beyond the late M dwarfs, lies the L dwarf sequence, which have $T_{\rm eff}$ ranging from 2250 K to 1400 K \citep{kirkpatrick1999}. While for earlier types there can be some hydrogen burning very low mass stars, most L dwarfs will be substellar-mass brown dwarfs. The sequence extends into the cooler T dwarf regime and to $T_{\rm eff}$ of $\sim$ 500 K \citep{lucas2010}. The Y dwarf spectral type, first proposed by \citet{kirkpatrick1999} has only recently seen its first confirmed objects with \citet{cushing2011} discovering six Y dwarfs from the Wide-field Infrared Survey Explorer (WISE). More recently \citet{kirkpatrick2012} presented seven more Y dwarfs from WISE, bringing the current total of confirmed objects to 13. The WISE \citep{wright2010} satellite has surveyed the entire sky in four infrared bands (\textit{W1}, \textit{W2}, \textit{W3} and \textit{W4}), at wavelengths of 3.4, 4.6, 12 and 22 $\mu$m. Bands \textit{W1} and \textit{W2} give the best sensitivity to the coolest brown dwarfs due to the strong absorption centred at 3.3 $\mu$m (\textit{W1}) owing to methane for objects with $T_{\rm eff}$ $<$ 1500 K. At 4.6 $\mu$m (\textit{W2}), however, there is no methane; as such the $\mathit{W1-W2}$ colour is very sensitive to cool brown dwarfs. 

Several other large-scale optical and near-infrared (NIR) surveys have already proved to be effective tools at finding and studying large populations of UCDs: the Two-Micron All-Sky Survey (2MASS; \citealt{skrutskie2006}) and the Deep Near Infrared Survey of the Southern Sky (DENIS; \citealt{epchtein1997}) in the NIR, and the Sloan Digital Sky Survey (SDSS; \citealt{york2000}) in the optical. As brown dwarfs cool and fade significantly over time this leads to a mass-age degeneracy, with lower mass younger brown dwarfs having similar $T_{\rm eff}$ and luminosity to higher mass, older ones. Measuring atmospheric properties ($T_{\rm eff}$, log \textit{g}, [m/H])  of brown dwarfs in order to infer mass and age, is thus crucial. Using near-infrared and optical spectra and model fitting techniques, one can attempt to constrain physical properties for UCDs. However dust condensation, non-equilibrium chemistry and complex molecular opacities make it very challenging to accurately model such atmospheres, and the reliable fitting of properties with spectral models is not currently reliable \citep{pinfield2012}. 

One way to overcome these problems is to identify UCDs whose properties can be estimated in a relatively  independent way. We refer to such objects as benchmark UCDs. They can be used as a testbed for prevailing theories and models. UCDs that are wide companions to main sequence (MS) stars of the Galactic disc are fairly numerous, cover the full range of age and metallicity, and are therefore particularly useful benchmarks. \citet{vanbiesbroeck1961} was one of the first to search for these systems and the method used has been followed by many other successfull studies (e.g. \citealt{kirkpatrick2001}; \citealt{wilson2001}; \citealt{gizis2001}; \citealt{pinfield2006}; \citealt{burningham2009}; \citealt{faherty2009, faherty2010}; \citealt{day-jones2011}; \citealt{pinfield2012}). 

Since the peak of the spectral energy distribution (SED) of L dwarfs occurs at NIR wavelengths, 2MASS and WISE are ideal to discover large numbers of these objects. In fact both surveys cover the whole sky and are well matched which makes them a powerful tool to detect L dwarfs. Despite their relatively large pixel sizes of 1 arcsec in 2MASS and 1.375 arcsec in WISE, the 10 year baseline between these surveys also makes it possible to accurately determine proper motions. This is important as it allows confirmation of the binarity in these systems.

We present here the measurement of proper motion for 10 L dwarf candidates and the discovery of two new wide binary systems. Section 2 describes the technique used to identify these systems and how we assessed their common proper motion, describing our follow-up spectroscopic observations whilst in section 3 we present the two new binary systems in more detail, addressing the issue of possible contamination. In section 4 we discuss results, estimate a binary fraction for these type of objects and investigate the possibility that some known UCDs with MS stars companions are actually unresolved objects. Finally our conclusions are presented in section 5.  

\section{Candidates selection}
\subsection{L dwarfs selection}

The UCD candidates were selected using NIR photometry available from the 2MASS all-sky point catalogue. This database is ideal to search for L dwarfs as it covers the whole sky and the NIR \textit{JHKs} filters sample the peak of the SED of such objects. Colour selection criteria were applied to obtain a first sample of L dwarf candidates. This selection required the following:
 
\begin{center}

0.5	$\leq$ $\mathit{J-H}$ $\leq$ 1.6 \\
1.1	$\leq$ $\mathit{J-K}$ $\leq$ 2.8 \\
0.4	$\leq$ $\mathit{H-K}$	$\leq$ 1.1 \\
$\mathit{J-H}$ $\leq$ 1.75 ($\mathit{H-K}$) + 0.37 \\
$\mathit{J-H}$ $\geq$ 1.65 ($\mathit{H-K}$) $\mathit{-}$ 0.35
\end{center}

We only consider candidates with a \textit{J} band magnitude $\leq$ 16.0. These constraints were chosen based on the colours of known L dwarfs from Dwarf Archives\footnote[1]{See http://DwarfArchives.org}. To avoid source confusion and contamination from MS and giant stars some high density regions were excluded, such as the galactic plane (b $\geq$ $\vert25^\circ \vert$) and the Large and Small Magellanic Clouds. Two other reddened and overcrowded regions were also removed. Information about these regions can be seen in Table \ref{tab:red_regions}. The equatorial poles were also avoided ($\delta$ $\geq \vert86^\circ\vert$) as 2MASS is not complete in these regions for optical cross-matching. The total area of the sky included in the first sample was 22, 178  deg$^2$, or 53 per cent of the sky. 
\indent

We have kept objects with no optical counterpart within 5 arcsec in the USNO-A2.0 catalogue or, if they do have counterparts, with $\mathit{R-K}$ $>$ 5.5. By doing this we ensure that all detections with optical counterparts are of spectral types later than M6 as these objects have extremely red colours in both the optical and IR domain due to their low $T_{\rm eff}$. A signal-to-noise ratio (S/N) criteria was imposed to exclude low-quality photometric data (S/N $>$ 5 for all the bands). Flags including ccflg=000, prox$>$6, gal$\_$contam=0 and mpflg=0, were applied to make sure that the sources were unaffected by known artefacts, such as diffraction spikes from nearby bright stars, and to exclude extended sources and known minor planets. This first sample of possible UCD candidates encompasses 28, 023 objects. 

\begin{table}
\caption{Uncatalogued regions that have been removed from the initial sample}
\label{tab:red_regions}
\begin{center}
\begin{tabular}{cccc}
\hline
\hline

l$_{min}$ & l$_{max}$ & b$_{min}$ & b$_{max}$ \\
\hline
0   & 96  & -16 & 16   \\
150 & 180 & 10  & 13   \\
180 & 360 & -13 & 13   \\
199 & 214 & -13 & -27  \\
308 & 310 & 13  & 16   \\
\hline
\end{tabular}
\end{center}
\end{table}

We find that 269 of the 602 known L dwarfs were present in our initial list. The remaining ones did not make it through our initial cuts, either because they had \textit{J} band magnitudes larger than 16, galactic latitudes outside our limit or they did not pass the quality flags imposed. It is important to note that only 36 were excluded based on their colours, thus, the vast majority of L dwarfs have NIR colours inside the colour space we initially defined, thus validating our method.

\subsection{Selection of binaries}

\subsubsection{Assessing photometry and separation}

In order to search for companions to our L dwarf candidates, we selected a sample of MS stars from different catalogues, namely the Hipparcos \citep{vanleeuwen2007} and Gliese \citep{gliese1991} catalogues. We selected only Hipparcos stars within 50 pc as we only want to consider pairs up to this distance, whilst Gliese stars have a maximum distance of 25 pc (due to limitations of the catalogue itself). The sample consists only of F, G, and K stars. Assuming the star's distance, we looked for L dwarf companions up to an on-sky separation corresponding to 20, 000 AU, and angular separations of up to 10 arcmin, finding 572 possible pairs. 
  
\begin{figure}
\begin{center}

\includegraphics[height=6cm,width=8.0cm,angle=0]{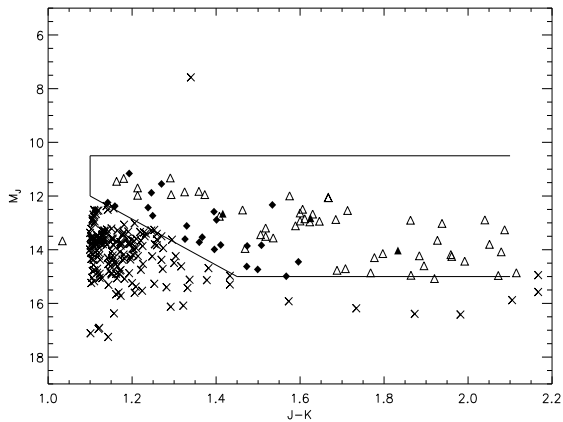}

\caption{$\mathit{M_J}$ versus $\mathit{J-K}$ colour-magnitude diagrams. The UCDs from the candidate pairs are shown as crosses. Inside the selection box, the filled triangles are the three known L dwarfs that turn up in our sample as possible binaries, the filled diamonds are UCDs candidates not confirmed and not in the dwarfarchive.org. The open triangles represent other UCDs with known parallax measurements from Dwarf Archives and are displayed for comparison purposes only.}
\label{fig:color_mag}
\end{center}
\end{figure}

Next, we plot the UCD candidates on a colour-magnitude diagram (CMD), in order to assess if the their photometry was consistent with a real UCD at the companion's distance. Using the distance of the primary star, we calculated $\mathit{M}_J$ and with the $\mathit{J-K}$ colour, obtained the plot shown in Fig. \ref{fig:color_mag}. Following the method used by \citet{pinfield2006}, we selected pairs where the L dwarf candidate was within a specific region. This region was defined using a sample of L dwarfs with known parallax, or with distances inferred from companion stars \citep{kirkpatrick2001,wilson2001} and the $\mathit{M_J}$ range from \citet{knapp2004}. We considered all of our pairs and using the CMD, identified 25 possible MS star-UCD binary systems.

\subsubsection{Cross-checking in other surveys}

In order to account for previous analysis, look for multi-epoch imaging data and additional colour constraints for the UCD candidates, we performed a cross-check with multiple NIR and optical surveys, specifically the SuperCOSMOS Science Archive (SSA), SDSS, Digitized Sky Survey (DSS), DENIS, WISE, the UKIRT Infrared Deep Sky Survey (UKIDSS) and the SIMBAD database. We excluded nine of the selected UCD candidates, as they were classified as galaxies or flare stars. Of the remaining 16 candidates, three were found to be known L dwarfs, another three have measured proper motions in the literature, and the remaining 10 do not have any proper motion measurements. This cross-check also allowed us to select the best epochs to calculate proper motions for the latter 10 UCD candidates. 
We used 2MASS as the first epoch and WISE as the second epoch when possible. Being the most recent NIR survey, WISE allows us to have a larger baseline between the two epochs and thus more accurate proper motion measurements. If WISE images were not the most suitable ones, we then selected the second epoch from the other surveys. All images were chosen taking into account the largest baselines possible and a reasonable number of appropriately positioned bright stars that could be used as reference objects. 

\subsubsection{Assessing common proper motion}

Proper motion calculations were made using standard IRAF routines. Measuring the positions of an average of 12 reference stars, we have used the IRAF routine GEOMAP to derive the spatial transformation between the two epochs. Then the routine GEOXYTRAN applied these transformations to the coordinates of the UCD candidate, and estimated how they changed between epochs. The uncertainties for this process are the combination of the \textit{rms} scatter associated with the coordinates transformation and the centroiding accuracies. Amongst our UCD candidates some had proper motions available in the literature. We compared our results with the published ones to check for consistency and then use the value with the smaller uncertainty. 

To verify if a pair has a common proper motion, we compared the proper motion in right ascension and declination of each component, taking into account the uncertainties associated with each measurement. We consider that two objects have common proper motion if they match to within $\sigma$ $<$ 2, $\sigma$ being the difference between their individual proper motions across the sky divided by the errors. In Table \ref{tab:proper_motion} we present the proper motions for all the candidates. Out of this sample, only two were found to have common proper motion with their primary star. The UCDs in these pairs are 2MASSJ133245.30+745944.1 and 2MASSJ130541.07+204639.4 (hereafter 2MASSJ1332+7459 and 2MASSJ1305+2046 respectively). Information about these is shown in Tables \ref{tab:G255-34} and \ref{tab:GJ499}. In one of the cases the UCD is an uncatalogued L dwarf, whilst the other is classified as an L4 $\pm$ 2 by \citet{cruz2003}. Our sample also  included one known binary discovered by \citet{faherty2010}, G 62-33.  

\begin{table*}
\caption{Proper motion measurements for the L dwarf candidates}
\label{tab:proper_motion}
\begin{tabular}{@{}ccccccccc}
\hline
\hline

 Name of L dwarf candidates & R.A. (J2000) & Dec (J2000) & pm$_{R.A.}$ & pm$_{Dec}$ & Possible primary & Notes \\
\hline
2MASSJ003142.93-630818.6 & 00 31 42.94 & -63 08 18.6 & 133 $\pm$ 8   & -88 $\pm$ 8  & HIP2540 & a \\
2MASSJ071051.38-492212.2 & 07 10 51.38 & -49 22 12.4 & -14 $\pm$ 12  & 121 $\pm$ 9  & HIP34739  & a \\
2MASSJ074231.27+180816.8 & 07 42 31.27 & +18 08 16.8 & 57 $\pm$ 31  & -78 $\pm$ 15 & HIP37622  & a\\ 
2MASSJ100428.24-114648.9 & 10 04 28.25 & -11 46 49.1 & -130 $\pm$ 15 & 91 $\pm$ 15  & HIP49366  & a \\
2MASSJ132434.95+545615.2 & 13 24 34.94 & +54 56 15.4 & 2 $\pm$ 12  & 7 $\pm$ 10  & HR5055 & a  \\
2MASSJ134154.14-014553.1 &13 41 54.15  & -01 45 53.2 & -18 $\pm$ 5 & -34 $\pm$ 6  & HIP66886 & a \\
2MASSJ134751.79-104433.7 & 13 47 51.79 & -10 44 33.7 & -432 $\pm$ 28 & -9 $\pm$ 18  & HIP67344 & a  \\
2MASSJ190536.28-370546.3 & 19 05 36.29 & -37 05 46.3 & -0.2 $\pm$ 2 & -7 $\pm$ 6 & HR7227 & a \\ 
2MASSJ073523.28+315050.6 & 07 35 23.18 & +31 50 50.6 & 14 $\pm$ 10 & 30 $\pm$ 8 & HD60179C & a \\
\hline
2MASSJ133245.30+745944.1 & 13 32 45.31 & +74 59 44.2 & -471 $\pm$ 28 & 39 $\pm$ 22 & G255-34 & b \\
2MASSJ130541.07+204639.4 & 13 05 41.07 & +20 46 39.4 & -23 $\pm$ 17 & 73 $\pm$ 27  & GJ499AB & b \\
\hline
2MASSJ083204.51-012836.0 & 08 32 04.51 & -01 28 36.1 & 64 $\pm$ 13  & 27 $\pm$ 15  & GJ3504 & c \\
2MASSJ132044.27+040904.5 & 13 20 44.28 & +04 09 04.7 & -483 $\pm$ 19 & 211 $\pm$ 17 & HIP65121  & d \\
2MASSJ112149.24-131308.4 & 11 21 49.25 & -13 13 08.4 & -509 $\pm$ 10 & -81 $\pm$ 10 & GJ3655 & e \\   
2MASSJ022128.59-683140.0 & 02 21 28.61 & -68 31 40.1 & 46 $\pm$ 6	& -6 $\pm$ 17 & HIP11001  & f \\
2MASS063447.73-582955.3  & 06 34 47.73 & -58 29 55.3 & 73 $\pm$ 9 & -21 $\pm$ 9 & HIP31300 
& g \\
\hline
\end{tabular}

\textbf{Notes:} (a) New L dwarf candidates with new proper motion measurements (b) new common proper motion pairs, here presented (c) proper motion from \citet{casewell2008} (d) proper motion from \citet{jameson2008} and common proper motion pair discovered by \citet{faherty2010} (e) proper motion from \citet{salim2003} (f) proper motion from \citet{faherty2009} (g) proper motion from \citet{roeser2010} 

\end{table*}

\subsection{Follow up observations}

The 2MASSJ1332+7459 UCD candidate was observed with the 3.58-m Telescopio Nazionale Galileo (\textsc{TNG}) in combination with the Device Optimized for the LOw RESolution (\textsc{dolores} or \textsc{lrs} for short) on 2011 May 13. \textsc{lrs} was equipped with a 2048 $\times$ 2048 E2V 4240 \textsc{ccd} and the \textsc{lr-r} grating with a resolving power of $R \sim 700$,  centred at 7400 \AA. A slit width of 1.0 arcsec was used with a seeing of $\sim 1.1$ arcsec during the observations. This setup provided a wavelength coverage from 5200 \AA\, to 10400 \AA, with a dispersion of 2.7 \AA\, per pixel and a resolution of 3.8 \AA\, at the central wavelength.  

The spectra were reduced in the following way: the average bias and flat field correction was carried out using the \textsc{figaro} package from \textsc{starlink} using average bias and halogen frames, taken on the same night. The spectral reduction suite \textsc{Pamela} was used for the optimal extraction of the spectra. 

A neon and mercury arc lamp exposure at the same position as the targets allowed us to establish an accurate wavelength scale for each of the five spectra obtained. We used \textsc{Molly}, a 1-D astronomical spectral analysis package, to fit the arc frame with a fourth order polynomial, giving a \textit{rms} of 0.17 \AA. The  spectral resolution was measured by the full width at half-maximum (\textsc{fwhm}) of the arc lines, using an average of 10 lines located across the entire wavelength range. The individual spectra were flux calibrated using the nearby spectro-photometric standard flux star \textsc{grw} +70d5824. The final spectrum is a variance weighted average of the five individual spectra.

\section{Binary candidates}

\subsection{G255-34AB}

\subsubsection{Properties of the L dwarf G255-34B}

\begin{figure}
\begin{center}

\includegraphics[height=7.5cm,width=8.3cm,angle=0]{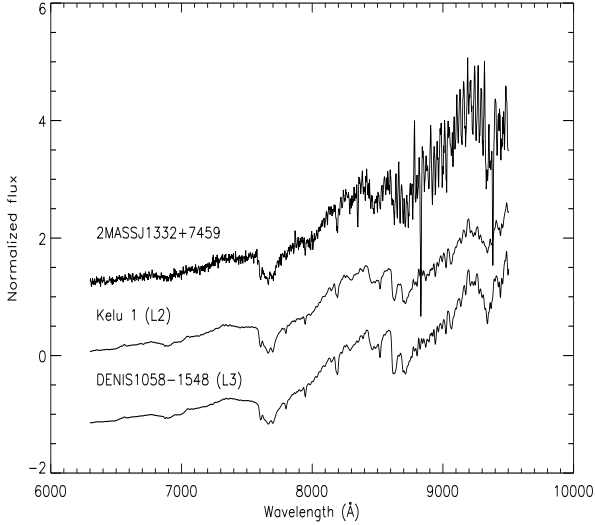}

\caption{Optical spectrum of the 2MASSJ1332+7459 dwarf, with the standard spectra of an L2 dwarf, Kelu 1, and an L3 dwarf, DENIS 1058-1548. The spectra has been displaced vertically in order to better compare them. The fit has been made for the wavelength range 6300-9500 $\mu$m.}

\label{fig:spectrum}
\end{center}
\end{figure}

To measure the UCDs spectral type, we have compared the optical spectra obtained at the TNG to standard template spectra from the L dwarf sequence defined in \citet{kirkpatrick1999}. The comparison was done using four different subclasses, L0, L1, L2 and L3, using the following objects as standards: for L0, 2MASP J034532+254023; for L1, 2MASSW J1439284+192915, for L2, Kelu1 (all three from \citealt{kirkpatrick1999}), and for L3, DENIS-P J1058.7-1548 \citep{delfosse1997}. In order to define the best subclass for this dwarf, we used a $\chi^2$ minimization and the goodness-to-fit statistic, \textit{G}$_k$, defined in \citet{cushing2008}. Whereas the $\chi^2$ minimization classifies this UCD as an L2 dwarf, the method described in \citet{cushing2008} classifies it as an L1.  The fit of the observed spectrum to the template spectra can be seen in Fig. \ref{fig:spectrum}. The fitting was done for the wavelength range 6300-9500 $\mu$m.

Secondly, we estimated three spectral ratios, CrH-a, Rb-b/TiO-b and Cs-a/VO-b, defined in \citet{kirkpatrick1999} for L dwarfs with spectral types $<$ L5. The values for these ratios are 1.29, 1.08 and 0.91 respectively, usually associated with L2 dwarfs. 

Looking more closely at the spectrum of G255-34B, we see that despite showing TiO $\lambda$8432, CrH $\lambda$8611 and FeH $\lambda$8692 lines all similar in strength, typical of L1 dwarfs, the spectrum actually shows most of the spectral key features of the subclass L2 \citep{kirkpatrick1999}. As an example, we refer to the distinctly sloped part of the spectrum between 7800 and 8000 \AA\ and the TiO lines $\lambda$8432 and $\lambda$7053 that become weaker and disappear, respectively.

Finally, we checked the two WISE bands to derive a spectral type in an independent way, using colour-spectral type plots. According to the $\mathit{W1-W2}$ vs spectral type plot of \citet{kirkpatrick2011}, the L dwarf, showing a $\mathit{W1-W2}$ colour of 0.254, should have a spectral type between M5 and L4. However, looking at the colours, $\mathit{J-W2}$ and $\mathit{H-W2}$ vs spectral type plots, it can be seen that the UCD colours are similar to the ones shown by L0 to L3 dwarfs. These conclusions are consistent with the value obtained using our previous method. Taking all these spectral classification diagnostics into account, we here classify this UCD as an L2 dwarf.    

To determine the bolometric luminosity we have combined the optical spectra of the L dwarf with the photometry available. In the NIR part of the spectrum we used \textit{JHK} photometry from 2MASS, extending to the mid infrared with WISE bandpasses \textit{W1} and \textit{W2}. We followed the method described in \citet{marocco2010} to estimate the bolometric luminosity. Firstly, we created synthetic spectra for those regions of the electromagnetic spectrum lacking observations (NIR and mid-IR). We used the models of \citet{hubeny2007} since these cover a temperature range between 700 to 1900 K and have considered that the models are still valid for the typically larger temperature values of early L dwarfs (up to 2250 K). We allowed for different values of log \textit{g}, 4.5, 5.0 and 5.5, and assumed K$_{zz}$ = 10$^{2}$, 10$^{4}$ and 10$^{6}$ cm$^{-2}$s$^{-1}$ (eddy diffusion coefficient). Secondly, we estimated the difference in flux between our models and the flux given by the available photometry, known as the scaling factor. Comparing this value with the scaling factor given by the distance assumed for the L dwarf and the radius range adopted (between 0.8 R$_{Jup}$ and 1.2 R$_{Jup}$) we only considered models for which the average scaling factor was within a certain interval. Thirdly, combining the synthetic spectra with the observed one, we calculated the bolometric flux, luminosity, and temperature range from the radius range. We then excluded the models in which the temperature was at least 100 K outside the constrained range. The final step takes into account only the remaining models and using the mean flux, we have estimated a final bolometric luminosity of 6.38 $\pm$ 0.67 $\times$ 10$^{29}$ erg s$^{-1}$ or 1.67 $\pm$ 0.17 $\times$ 10$^{-4}$ L$_{\odot}$. Taking a radius range of 0.8-1.2 $\mathit{R_{Jup}}$ and the luminosity, we can derive the corresponding temperature for this L2 dwarf as $T_{\rm eff}$ = 2080 $\pm$ 260 K.   

We have also derived the bolometric luminosity and $T_{\rm eff}$ for the L dwarf following this procedure: taking into account the spectral type information derived before, we estimated the \textit{K} band bolometric corrections using the polynomial fit of \citet{golimowski2004} and hence the apparent bolometric magnitude of the object. Combining this with the distance, inferred from parallax measurements, we obtain the absolute bolometric magnitude and consequently, the bolometric luminosity. From the luminosity, $T_{\rm eff}$ can be directly calculated. The results are, for bolometric luminosity and $T_{\rm eff}$ respectively, 7.14 $\times$ 10$^{29}$ erg s$^{-1}$ or 1.87 $\times$ 10$^{-4}$ L$_{\odot}$, and $T_{\rm eff}$ = 2106$^{+250}_{-180}$ K, and are in agreement with the previous values. For the $T_{\rm eff}$ we again considered a radius range of 0.8-1.2 $\mathit{R_{Jup}}$.

\subsubsection{System properties}

\begin{figure*}
\begin{center}

\includegraphics[height=6.5cm,width=13.0cm,angle=0]{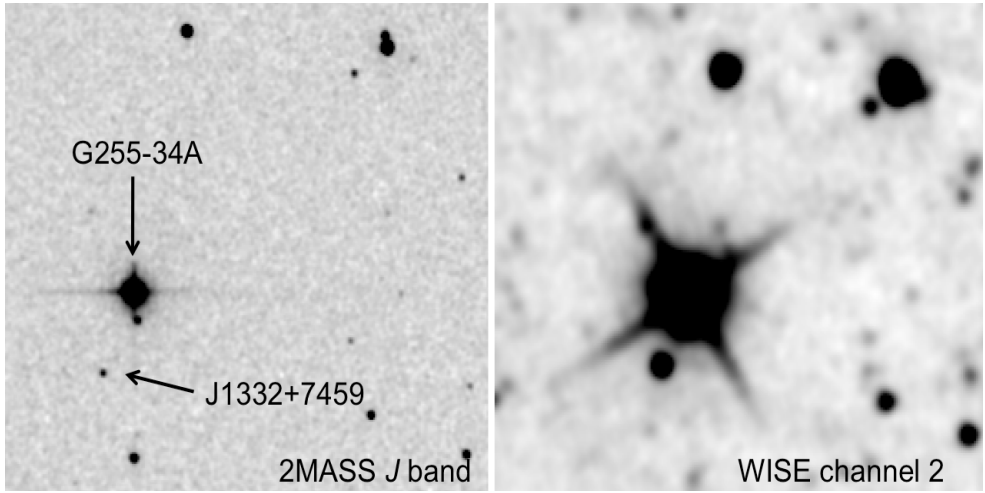}

\caption[fig:chart]{A 4.5 arcmin $\times$ 4.5 arcmin finder chart for the system G255-34AB. On the left panel is the \textit{J} band image and on the right one the \textit{W}2 band. North is up and East is left.}

\label{fig:chart}
\end{center}
\end{figure*}

The components of the first system are the L2 dwarf discussed in Section 3.1.1 and a K8 dwarf \citep{vyssotsky1956}, G255-34A, and a finder chart of these can be seen in Fig. \ref{fig:chart}. The spectral type we derived earlier for the L dwarf suggests an $\mathit{M_J}$ of 12.36 $\pm$ 0.11 based on \citet{dupuy2012}. This places it at a distance of 35.36 $^{+1.84}_{-1.75}$ pc, consistent with the measured parallax of the K8 primary, at 36.21 $\pm$ 1.50 pc \citep{vanleeuwen2007}.
G255-34A, a K8 dwarf, has a separation of 38.3 arcsec to the secondary, or 1364 AU. Initially believed to be part of a common proper motion pair with another star by \citet{luyten1979}, it is not considered as such by \citet{giclas1971}, as can be noted in  \citet{weis1991}.

\citet{schlaufman2010} have developed a way to calibrate M dwarf metallicity using photometry. Applying their method and using \textit{V} and \textit{Ks} magnitudes, we find [Fe/H] to be $-$0.29 dex. However, metallicity uncertainties are significant for photometric constraints, $\sim$ 0.2 dex. This is marginal evidence for G255-34A being slightly metal poor. 
Kinematics place this star outside the region defined as the Eggen box \citep{eggen1989} (and representative of the young disc population) and therefore suggest old disc membership and a likely age greater than 1.5 Gyr. According to \citet{leggett1992} this is broadly consistent with an [m/H] $\sim$ $-$0.5 and we employ the relationship for West's earliest spectral type range as reasonably representative for a K8 dwarf to place a lower limit on the age of this system at 0.8$\pm$0.6 Gyr. 

\begin{table}
\caption{Properties of the system G255-34AB}
\label{tab:G255-34}
\begin{center}
\begin{tabular}{cc}
\hline
Parameter & Value \\
\hline
Name & 2MASSJ1332+7459 (G255-34B)\\
R.A. (J2000) & 13 32 45.31  \\
Dec (J2000) & +74 59 44.2 \\
\textit{J}  & 15.103 $\pm$ 0.041 \\
\textit{H}  & 14.116 $\pm$ 0.034 \\
\textit{K}  & 13.569 $\pm$ 0.032 \\
\textit{W1} & 13.170 $\pm$ 0.025 \\
\textit{W2} & 12.916 $\pm$ 0.032 \\
\textit{W1} -- \textit{W2} & 0.254 \\
\textit{J} -- \textit{W2} & 2.187 \\
\textit{H} -- \textit{W2} & 1.200 \\
$\mu_{\rm R.A.}$ & -471.296 $\pm$ 28.489 mas yr $^{-1}$\\
$\mu_{\rm Dec}$ & 38.773 $\pm$ 22.113 mas yr $^{-1}$\\ 
SpType & L2 \\
log $\mathit{L/L_\odot}$ & -3.780 $\pm$ 0.045 \\ 
$T_{\rm eff}$ & 2080 $\pm$ 260 K \\ 
\hline
Name & G255-34A (HIP66074) 	 \\
R.A. (J2000) & 13 32 41.02  \\
Dec (J2000) &  +75 00 24.81\\
\textit{B} & 11.53 \\
\textit{V} & 10.23 \\
\textit{R} & 9.7 \\
\textit{I} & 9.2 \\
\textit{J} & 7.910 $\pm$ 0.024 \\
\textit{H} & 7.302 $\pm$ 0.033 \\
\textit{K} & 7.182 $\pm$ 0.016 \\
$\mu_{\rm R.A.}$ & -438.50 $\pm$ 1.33 mas yr $^{-1}$\\
$\mu_{\rm Dec}$ &  49.82 $\pm$ 1.05 mas yr $^{-1}$\\ 
Parallax & 27.62 $\pm$ 1.14 mas \\
Distance & 36.21 $\pm$ 1.50 pc \\
SpType & K8 \\
U & -49.7 km s$^{-1}$ \\   
V & -63.8 km s$^{-1}$ \\
W & -17.6 km s$^{-1}$ \\
V total & 82.7 km s$^{-1}$ \\
$T_{\rm eff}$ & 4368 K\\
\hline
\end{tabular}
\end{center}
\end{table}

\subsection{GJ499ABC}

\subsubsection{Properties of the L dwarf GJ499C}

GJ499C is an L dwarf first discovered by \citet{cruz2003}. The proper motion for this dwarf has been estimated by some authors and the available values for this are shown in Table \ref{tab:proper_motion_gj499c}. Our own measurements are also presented and, within the error bars, are in good agreement with the values of \citet{jameson2008}, \citet{faherty2009} and \citet{schmidt2010}. 

\begin{figure*}
\begin{center}

\includegraphics[height=6.5cm,width=13.0cm,angle=0]{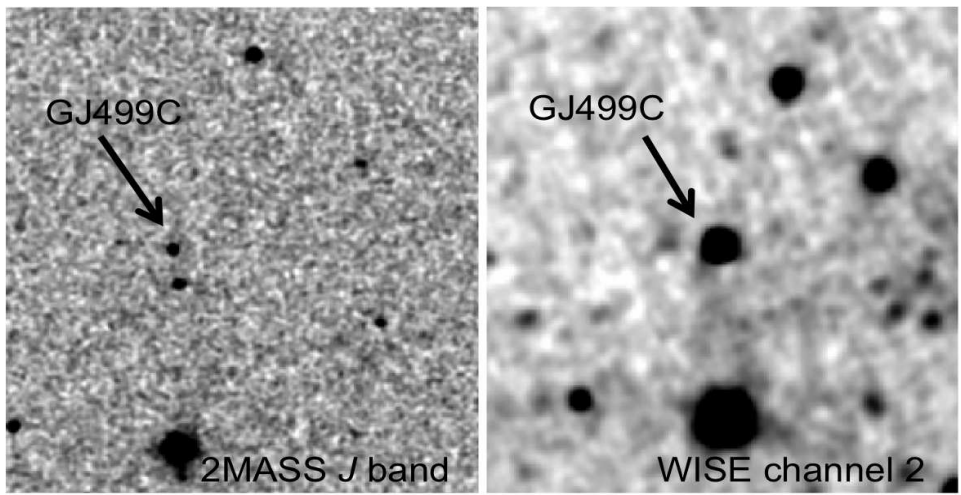}

\caption[fig:chart2]{A 5 arcmin $\times$ 5 arcmin finder chart for the GJ499C dwarf. On the left panel is the \textit{J} band image and on the right one the \textit{W}2 band. North is up and East is left.}

\label{fig:chart2}
\end{center}
\end{figure*}

We have independently estimated the spectral type of the dwarf with the available spectrum from SDSS. We started by calculating the CrH-a, Rb-b/TiO-b and Cs-a/VO-b spectral ratios, with the additional colour-d ratio (since these are best suited to objects with spectral types $>$ L5). The analysis of spectral ratios are all consistent with a spectral type of L5. We then compared the SDSS spectra with standard spectra templates, following the $\chi^2$ minimization and the G$_k$ factor method. We did this in order to assess if the fit to the spectral templates was consistent with the spectral type indicated by the spectral ratios alone. We find that $\chi^2$ minimization classifies the L dwarf as an L5, whereas the G$_k$ factor suggests an L4 type. A final spectral type of L5 is adopted, which is in agreement with the previous result by \citet{cruz2003}, that classifies it as an L4 $\pm$ 2. A finder chart for this L dwarf can be seen in Fig. \ref{fig:chart2}.

Following the method described in section 3.1.1, we have estimated the luminosity and $T_{\rm eff}$ for GJ499C, using bolometric corrections and a radius range of 0.8-1.2 $\mathit{R_{Jup}}$. The results are, for the luminosity and $T_{\rm eff}$, 2.23 $\times$ 10$^{29}$ erg s$^{-1}$ or 5.83 $\times$ 10$^{-5}$ L$_{\odot}$, and 1574$^{+190}_{-140}$ K, in agreement with what it is expected for L5 dwarfs.

\subsubsection{System properties}

The primary components of the system G499AB contain a K5 with a closely separated M4 dwarf \citep{reid2004}, at a separation of 0.9 arcsec. According to \citet{dupuy2012}, the mean $\mathit{M_J}$ value for a L5 dwarf is 13.56 $\pm$ 0.03, and with this we estimate a distance of 21.28$^{+0.30}_{-0.28}$ pc to the GJ499 triple system. The parallax measurements from the Hipparcos catalogue \citep{vanleeuwen2007} show that the primary is at a distance of 18.80 $\pm$ 0.61 pc, and is in agreement with the photometric distance derived for the UCD companion. GJ499C has a on-sky separation of 8.6 arcmin from the primary, or 9708 AU. Proper motion measurements come from the \citet{vanleeuwen2007} catalog and are in agreement with other values published such as \citet{zacharias2012}, \citet{kharchenko2009} and \citet{roeser2008}.
In order to estimate the physical properties of this system, we follow the same approach as for G255-34AB. 

According to the method described in \citet{schlaufman2010} we can use the \textit{V} and \textit{Ks} magnitudes of 9.439 and 6.041 respectively (taken from the Hipparcos and 2MASS catalogues) for the K5 dwarf to estimate a [Fe/H] = $-$0.20 $\pm$ 0.2, suggestive of a slightly metal poor system. If we do the same calculations for the M4 companion, with \textit{V} = 14.90 and \textit{Ks} = 9.55, the result is  [Fe/H] = $-$0.17 $\pm$ 0.2. Both results are consistent with a metal poor system. 

The velocity components published in \citet{bobylev2006} place the pair on the edge of the Eggen box, therefore suggesting youth. However, kinematics offers a poor tool to place constraints on ages. \citet{west2008} used the activity lifetimes of M dwarfs to calculate ages, suggesting an age of 4.5$^{+0.5}_{-1.0}$ Gyr for M4 dwarfs. We thus consider this as a lower limit to the age of the system.

\begin{table}
\caption{Properties of the system GJ 499ABC}
\label{tab:GJ499}
\begin{center}
\begin{tabular}{cc}
\hline
Parameter & Value \\
\hline
Name & J1305+2046 (GJ 499C)\\
R.A. (J2000) & 13 05 41.07  \\
Dec (J2000) & +20 46 39.4 \\
\textit{J}  & 15.20 $\pm$ 0.053 \\
\textit{H}  & 14.04 $\pm$ 0.042 \\
\textit{K}  & 13.37 $\pm$ 0.039 \\
\textit{W1} &  12.540 $\pm$ 0.026 \\
\textit{W2} &  12.153 $\pm$ 0.025  \\
\textit{W1} -- \textit{W2} &  0.387 \\
\textit{J} -- \textit{W2} & 3.047 \\
\textit{H} -- \textit{W2} & 1.887 \\
$\mu_{\rm R.A.}$ & -42.61 $\pm$ 32.1 mas yr $^{-1}$\\
$\mu_{\rm Dec}$ & 103.25 $\pm$ 15.4 mas yr $^{-1}$\\ 
SpType & L5 \\
log $\mathit{L/L_\odot}$ & -4.23 $\pm$ 0.05 \\ 
$T_{\rm eff}$ & 1574 $\pm$ 170 K \\ 
\hline
Name & GJ 499AB (HIP63942) \\
R.A. (J2000) & 13 06 15.43  \\
Dec (J2000) &  +20 43 44.40 \\
\textit{V$_1$} & 9.40 \\
\textit{J$_1$} & 6.89 $\pm$ 0.01 \\
\textit{H$_1$} & 6.27 $\pm$ 0.01\\
\textit{K$_1$} & 6.05 $\pm$ 0.01 \\
\textit{V$_2$} & 14.90 \\
\textit{J$_2$} & 10.25 $\pm$ 0.10 \\
\textit{H$_2$} & 9.75 $\pm$ 0.10  \\
\textit{K$_2$} & 9.55 $\pm$ 0.10  \\
$\mu_{\rm R.A.}$ & -49.17 $\pm$ 1.7 mas yr $^{-1}$\\
$\mu_{\rm Dec}$ &  101.14 $\pm$ 1.1 mas yr $^{-1}$\\ 
Parallax & 53.18 $\pm$ 1.73 mas \\
Distance & 18.80 $\pm$ 0.61 pc \\
SpTypeA & K5\\
Sp TypeB & M4 \\ 
$T_{\rm eff}$ & 4210 K\\
U & -9.0 km s$^{-1}$ \\   
V & 5.1 km s$^{-1}$ \\
W & -1.4 km s$^{-1}$ \\
V total & 10.4 km s$^{-1}$ \\
\hline
\end{tabular}
\end{center}
\textbf{Notes:} \textit{V$_1$}, \textit{J$_1$}, \textit{H$_1$} and \textit{K$_1$} refer to GJ 499A whereas \textit{V$_2$}, \textit{J$_2$}, \textit{H$_2$} and \textit{K$_2$} refer to GJ 499B. 
\end{table}

\begin{table}
\caption{Proper motions for GJ 499C}
\label{tab:proper_motion_gj499c}
\begin{center}
\begin{tabular}{cccc}
\hline
\hline
Ref. & $\mu_{\rm R.A.}$ (mas yr$^{-1}$) & $\mu_{\rm Dec.}$ (mas yr$^{-1}$) \\ 
\hline
\citealt{zacharias2005} & -23.9 $\pm$ 9.0 & -73.1 $\pm$ 9.0 \\ 
\citealt{jameson2008}   & -23.3 $\pm$ 17.5 & 73.4 $\pm$ 26.9 \\
\citealt{faherty2009}   & -23.0 $\pm$ 17.0 & 73.0 $\pm$ 27.0 \\
\citealt{schmidt2010}   & -59.1 $\pm$ 33.1 & 73.8 $\pm$ 22.6 \\
This paper              & -42.6 $\pm$ 32.1 & 103.3 $\pm$ 15.4 \\
\hline
\end{tabular}
\end{center}
\end{table}

\subsection{Possible contamination}

We performed a statistical analysis with the purpose of estimating the probability that these two systems are, in fact, gravitationally bound and are not merely randomly aligned. 

Firstly we estimated the absolute magnitude in the \textit{J} band for each UCD $\pm$ 1 spectral type. We did this using two different methods, firstly the mean 2MASS $\mathit{M_J}$ as function of spectral type from the \citet{dupuy2012} and also from the \citet{marocco2010} relations between $\mathit{M_J}$ and spectral type. \citet{marocco2010} uses the MKO photometric system, which we converted the 2MASS photometry using the \citet{stephens2004} relations. With $\mathit{M_J}$ we estimate a minimum and maximum distance for each L dwarf. We then calculate a conical volume in the sky, with a radius that is defined by the separation between the UCD and its primary companion. For G255-34B the distances 24.5 pc and 45.8 pc define a volume of 2.95$\times 10^{-3}$ pc$^3$, whereas for the GJ499C, 12.6 pc and 27.0 pc define a volume of 0.116 pc$^3$. 

Secondly, we consider the number of stars within these volumes. The luminosity function \citep{reid2007} derived from the 8 pc and 20 pc samples gives us a lower and upper limit for the density of stars in the sky (0.062 to 0.076 stars pc$^{-3}$). Multiplying this number by the volumes around G255-34B, we expect to find between 1.830$\times 10^{-4}$ to 2.242$\times 10^{-4}$ MS stars that could appear as companions to the secondaries. For GJ499C we would expect between 7.185$\times 10^{-3}$ and 8.807$\times 10^{-3}$ stars.  

Another important aspect to consider is that, even though stars could lie within the volume around the L dwarfs, only a small fraction could have common proper motion. We therefore analyse the probability of finding Hipparcos and Gliese stars that, by random chance, may have the same motion across the sky as our L dwarfs. 
In order to do so, we used the initial sample of MS stars with a maximum distance of 50 pc and considered firstly the ones lying inside a spherical volume with 45$^{\circ}$ radius. Such a large volume assures that we have enough stars in our sample to estimate the fraction of those that could appear as common proper motion to our secondaries. We have carefully excluded from this region a circular area where the common proper motion primary companion lies, to avoid counting it as one of the randomly aligned stars. Next, we selected only stars that could masquerade as common proper motion companions to the L dwarfs up to a 2$\sigma$ detection, which was used in our initial classification as a common proper motion candidate, i.e., within $\pm$ 120 mas yr$^{-1}$ of the value of the secondary (valid for the two L dwarfs). 

Finally, taking into account the fraction of stars with similar proper motions as the secondary, and the sky density, we find that for the G255-34B dwarf, an average of 6.40$\times 10^{-6}$ MS stars could masquerade as a companion to the UCD by random chance, whereas for the GJ499C this value is of 1.36$\times 10^{-4}$ MS stars.  This statistical analysis allows us to say then, with a high degree of confidence, that both systems are genuine associations and not chance alignments to a high degree of significance.

\section{Discussion}

\begin{table*}
\caption{Known binary systems containing one L dwarf secondary}
\label{tab:ms_binaries}
\begin{center}
\begin{tabular}{@{}cccccccc}
\hline
\hline

Name & SpType & SpType & Distance (pc) & sep(arcsec) & projected sep(AU) & Age (Gyr) & Ref \\
             &  (primary) & (secondary) & (primary) & & & \\       
\hline

HD 89744A    & F8    & L0        & 39.43 $\pm$ 0.48 & 63    & 2480  & 2.1 - 7.2   & 4,13,25,32,33 \\
G 239-25A    & M3    & L0        & 10.73 $\pm$ 0.15 & 2.80  & 30    & -           & 4,17,25 \\
NLTT2274A    & M4    & L0        & 35.00 $\pm$ 2.00 & 23.0  & 483   & 4.5 - 10.0  & 23 \\
AB PicA      & K1    & L1        & 46.06 $\pm$ 1.46 & 5.5   & 253   & 0.02 - 13.8 & 4,8,25,26,31 \\
HD 16270A    & K3.5  & L1        & 21.27 $\pm$ 0.43 & 11.9  & 254   & -           & 4,7,25 \\
G 124-62A    & dM4.5 & L1+L1     & 27.48 $\pm$ 2.70 & 44    & 1452  & -           & 15,16,25 \\
GQ LupA      & K7    & L1.5      & 140.00 $\pm$ 50.00 & 0.7 & 98    & 3.0         & 20,23 \\
G 73-26A     & M2    & L2        & 35.00 $\pm$ 3.00 & 73.0  & 1898  & 3 - 4       & 23 \\
G 196-3A     & M3  & L2        & 14.90 $\pm$ 2.70 & 16    & 243   & 0.025       & 11,12,36  \\
GJ 618.1A    & K7 	 & L2.5	     & 33.42 $\pm$ 3.00 & 35    & 1170  & -           & 4,13  \\
G 62-33A     & K0    & L3        & 30.96 $\pm$ 0.82 & 66    & 2043  & 4.4         & 23,26 \\
G 200-28A    & G5    & L4        & 45.66 $\pm$ 1.29 & 570   & 2600  & 4.2 - 9.0   & 23,26,27 \\   
HD 49197A    & F5    & L4        & 44.90 $\pm$ 1.21 & 0.95  & 44    & 0.5 - 4.7   & 4,9,26,27,28 \\
GJ 564A      & F9    & L4+L4     & 18.17 $\pm$ 0.11 & 2.64  & 48    & 0.05 - 12.4 & 4,18,29,30 \\
HD 2057A     & F8    & L4+L4   & 43.82 $\pm$ 1.69 & 218   & 9465  & 1.6 - 5.3   & 4,5,6,23,26,34,37  \\
Gl 417A      & G0+G0 & L4.5+L4.5 & 21.93 $\pm$ 0.21 & 90    & 1953  & 0.08 - 3.2  & 14,15,27,29 \\
GJ 1001A     & M3.5  & L4.5+L4.5 & 13.01 $\pm$ 0.67 & 18.6  & 180   & -           & 1,2,3,25  \\
G 203-50A    & M4.5  & L5        & 22.40 $\pm$ 1.90 & 6.4   & 135   & -           & 21 \\
LP 261-75A   & M4.5  & L6        & 62.15 $\pm$ 28.58 & 13   & 450   & 0.04        & 10,12,25 \\
HD 203030A    & G8    & L7.5     & 40.88 $\pm$ 1.24 & 11.9 & 487   & 0.25 - 1.4  & 4,22,25,26,28 \\
Gl 584A      & G0+G3 & L8        & 17.86 $\pm$ 0.25 & 194   & 3465  & 3.3 - 4.8   & 19,26,34 \\
Gl 337A      & G8+K1 & L8+T0     & 20.36 $\pm$ 0.22 & 43    & 875   & 13.8        & 13,24,26 \\
\hline
G 255-34A      & K8    & L2       & 35.87 $\pm$ 149  & 38.0 & 1364  & -           & 38 \\
GJ 499AB      & K7+M4 & L5       & 18.80 $\pm$ 0.61 & 516 & 9708  & -           & 38 \\  
\hline

\end{tabular}
\end{center}

\textbf{References:} (1) \citealt{golimowski2004}; (2) \citealt{martin1999}; (3) \citealt{henry2006}; (4) \citealt{anderson2011}; (5) \citealt{liu2010}; (6) \citealt{cruz2007}; (7) \citealt{gizis2001}; (8) \citealt{chauvin2005}; (9) \citealt{metchev2004}; (10) \citealt{reid2006}; (11) \citealt{rebolo1998}; (12) \citealt{shkolnik2009}; (13) \citealt{wilson2001}; (14) \citealt{kirkpatrick2001};  (15) \citealt{bouy2003}; (16) \citealt{seifahrt2005}; (17) \citealt{forveille2004}; (18) \citealt{potter2002}; (19) \citealt{mugrauer2007}; (20) \citealt{neuhauser2005}; (21) \citealt{radigan2008}; (22) \citealt{metchev2006}; (23) \citealt{faherty2010}; (24) \citealt{burgasser2005}; (25) \citealt{dupuy2012}; (26) \citealt{casagrande2011}; (27) \citealt{holmberg2009}; (28) \citealt{wright2004}; (29) \citealt{lafreniere2007}; (30) \citealt{lambert2004}; (31) \citealt{tetzlaff2011} (32) \citealt{edvardsson1993}; (33) \citealt{bryden2009}; (34) \citealt{marsakov1995}; (35) \citealt{reid2003}; (36) \citealt{zapatero2010}; (37) \citealt{reid2006b}; (38) this paper

\end{table*}

\subsection{Binary fraction}
 
We have gathered information about all of the L dwarfs that are in binary or multiple systems, from the current literature, finding 59 L dwarfs that form part of binary or multiple systems. Of these, 35 are in a system where the primary is another UCD. Two have white dwarfs as primaries and the remaining 22 have a MS star primary. For the purposes of this paper, we only present the latter 22, with information about these systems shown in Table \ref{tab:ms_binaries}. The distance has been estimated using parallax measurements where available or, otherwise, from photometric data. 

Considering the two new systems presented here, the sample has now 24 binaries, with six triples and two quadruples. This gives a fraction of 1:4 triple to binary systems, and 1:12 quadruples to binary systems. These values differ from the ones presented in \citet{faherty2010}. However, in the latter case the authors considered a wide UCD companion sample containing all UCD binary systems, and not only the ones with L dwarfs as secondaries. 

Our initial sample of L dwarf candidates from 2MASS was completed up to a magnitude of 16 in the \textit{J} band. This is quite similar to the completeness limit of the Point Source catalogue of 2MASS. According to \citet{skrutskie2006}, 2MASS is virtually complete ($\sim$ 99 per cent of the sky) to the 10 $\sigma$ sensitivity limit for sources at or fainter than \textit{J} = 15.8. Our sample is thus complete to this limit and we see that all known L dwarfs with \textit{J} $< $ 16 have turned up in our initial search. Using the relationship between absolute and apparent magnitude (\citealt{marocco2010}, \citealt{knapp2004} and \citealt{liu2006}), for an L9 dwarf, we estimate that our sample is complete up to a distance of 19 pc (for an L0 this limit would be 94 pc). However, this completeness is only valid for the sky regions surveyed, since we have excluded reddened areas and avoided the galactic plane.

In order to estimate the number of L dwarfs expected in the full sky, we took a spherical region of space with a radius of 19 pc, and multiplied it by the \citet{cruz2007} space density estimate L dwarfs (3.8 $\pm$ 0.6 $\times$ 10$^{-3}$ pc$^{-3}$). We estimate that $\sim$ 109 $\pm$ 17 L dwarfs should lie within the region studied. 

Based on Table \ref{tab:ms_binaries}, we find that six L dwarfs are part of binary systems in which the primary is a MS star, including the GJ 499ABC system presented here. Therefore, we find that the binary fraction for L dwarfs that have MS companions is $\sim$ 6 per cent. Considering now all the Hipparcos and Gliese stars, we find that 1787 were within our area and out to 19 pc. The binary fraction of stars with L dwarfs as wide companions is thus 0.33 per cent. The systems considered for this analysis have separations between 30 and 10000 AU. It is worth noting that the two systems with smaller on-sky and physical separations, G 239-25AB and GJ 564AB (with separations of 30 and 48 AU respectively), have been discovered with adaptive optics, since finding such faint objects so close to bright stars is rather difficult.

If we now analyse the sample of known L dwarfs (based on Dwarf Archives), and do not take into account any distance constraints, we see that the binary fraction of L dwarfs in any kind of system goes up to $\sim$ 10 per cent. However, only 4 per cent of L dwarfs have MS stars as companions. We have not considered any possible unresolved binaries, and hence the values presented are just a lower estimate. The binary fractions here derived are not affected by the Malmquist bias, since both the number of single L dwarfs and the number of those found to be in multiple systems are affected in the same way.  

\subsection{Unresolved binaries}

\begin{figure*}
\begin{center}

\includegraphics[height=9cm,width=11.0cm,angle=0]{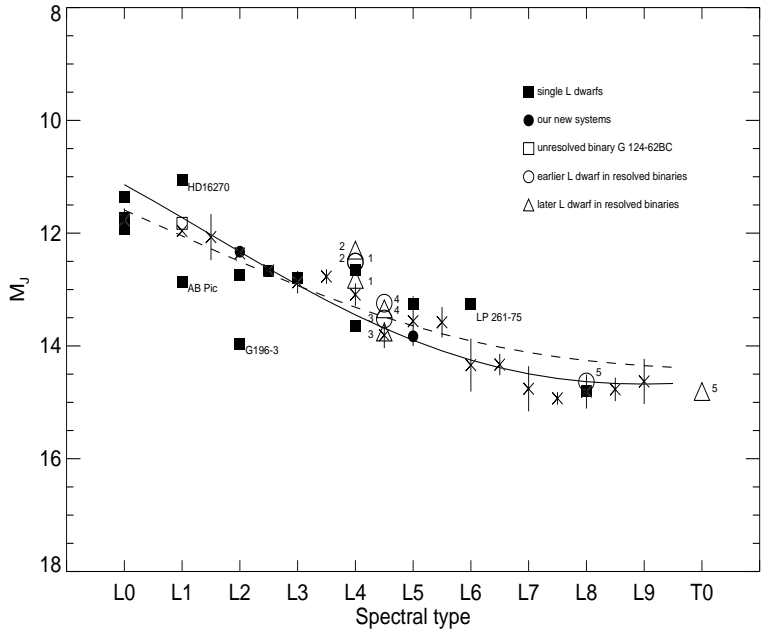}

\caption{Spectral type vs $\mathit{M_J}$ for the L dwarf + MS star systems in Table \ref{tab:ms_binaries}. The filled squares are those systems with only one L dwarf, the filled circles are the two new systems presented in this paper. The systems in which there are two unresolved L dwarfs and a MS star are represented by two symbols, an open circle showing the earlier of the two UCDs and an open triangle showing the later type UCD. The dashed line shows the polynomial fit from \citet{faherty2012} and the solid one the fit from \citet{marocco2010}. The crosses represent the mean 2MASS $\mathit{M_J}$ for each spectral type \citep{dupuy2012}. The numbers represent the following systems: (1) GJ 564ABC; (2) HD2057ABC; (3) Gl 417ABCD; (4) GJ 1001ABC; (5) Gl 337ABCD.}

\label{fig:abs_sptype}
\end{center}
\end{figure*}

In order to identify any possible unresolved binaries in the sample presented in Table \ref{tab:ms_binaries} we take advantage of the associated high quality parallax measurements that can be inferred for the companion L dwarfs in known multiple systems. We plotted the spectral type of the L dwarf components versus their $\mathit{M_J}$ (Fig. \ref{fig:abs_sptype}). To estimate the absolute magnitude we used the primaries distances. We excluded from this analysis two objects, GQ LupB (J15491209-3539039) and HD203030B (J21185897+2613461) since they do not have 2MASS \textit{J} band measurements. Also, we have plotted the G 124-62BC components as an unresolved binary, since there are no individual photometric measurements of the two L1 components. The other five systems in which a MS star has two UCDs as secondaries are all resolved, and in Fig. \ref{fig:abs_sptype} we plot the two components with different symbols, open circle and open triangle, and numbered. The two new benchmark systems here presented are shown as filled circles. The filled squares are L dwarf companions that have not previously been considered as possible unresolved multiples.

One interesting case is G 196-3B. The distance to this L2 dwarf has been estimated by many different authors. \citet{shkolnik2009} uses the companion, an M3 dwarf, to infer a photometric distance of 14.9 $\pm$ 2.7 pc to the system \citep{reid2002,reid2007}. \citet{faherty2009}, on the other hand, uses 2MASS $\mathit{M_J}$ vs spectral type relationships from \citet{cruz2003} to derive a different distance for this L2 dwarf, 32.0 $\pm$ 2.0 pc. More recently, \citet{zapatero2010} discuss the system in detail giving minimum and maximum values for its distance of 15 and 51 pc, respectively. These two limits are given for two different scenarios, one if we consider the primary to be a single object and a field star, and secondly if we consider the secondary to actually be a double object with ages of 3 Myr. Adopting a probable distance range of 15-30 pc (as \citealt{zapatero2010} conclude and in accordance with the \citealt{faherty2009} and \citealt{shkolnik2009} values), we estimate an $\mathit{M_{J}}$ between 12.4 and 13.9. In \ref{fig:abs_sptype} we have plotted the latter value, since this places the L2 dwarf further away from the polynomial fits. It can be seen that this particular dwarf shows redder colours than expected for its spectral type, as noted in \citet{zapatero2010} (this is true even if we consider the dwarf to be an L3, as \citealt{zapatero2010}). The reason for this is still to be fully explained, although \citet{zapatero2010} suggest the object might have a low gravity atmosphere with upper atmospheric layers or a warm dusty envelope.


The L dwarf companion to AB PicA has been previously reported as underluminous \citep{faherty2009}. This is a young system, with ages smaller than 1 Gyr, and a low gravity dwarf. We note though that the unresolved binary G 124-62BC, open square in Fig. \ref{fig:abs_sptype}, does not sit above the expected position in the spectral type vs $\mathit{M_J}$ plot. This could perhaps suggest that the actual single object sequence is slightly below the best-fitting sequences plotted, in which case AB PicB would not be underluminous, but making the position of HD16270B stand out even more as an overluminous object. Just like HD16270B, LP 261-75B appears to be slightly overluminous for its spectral type. However, if we take into account uncertainties in the spectral type of $\pm$ 1, we see that they both fit the sequence. 

Another interesting object is the system with HD2057A as primary, and two resolved L4 dwarfs as secondaries (number 2 in Fig. \ref{fig:abs_sptype}. Firstly mentioned in \citet{reid2006b}, the two L4 dwarfs are later confirmed as companions to the F8 star in \citet{faherty2010}. As can be seen in the plot, the two symbols sit well above the average $\mathit{M_J}$ for typical L4 dwarfs. In fact, one of the dwarfs has $\mathit{M_J}$ $\sim$ 1.15 above the value given by the polynomial fit of \citet{marocco2010} and 1.01 above the one by \citet{faherty2012}. This is not reported as a young system \citep{casagrande2011}, and therefore the overluminosity might be explained by multiplicity. In such case, we could be looking at a quadruple system of L dwarfs. Only with further observations can we expect to truly explain this system. 
In the same way, the GJ 564 system also appears to have one of the two L4 secondaries above the sequence, and again binarity could be the explanation for its position in the plot. 

\section{Conclusions}

Using the all-sky surveys 2MASS and WISE, we have identified two new systems with an L dwarf and a MS star as the primary components. Such benchmark systems allow us to better understand how the observed properties of UCDs are dependent on their ages and metallicities, and the two new systems here presented will help populate the growing sample of brown dwarfs with well constrained properties. Spectroscopy of the two primaries will be needed in the future to further estimate their ages and metallicities, thus fully characterizing the systems. 

We have also investigated the binary fraction for the specific type of binaries presented here, those in which the L dwarf has a FGK or M star as a wide companion. The number of such systems known to date is, however, small and any conclusions are dependent on the completeness of the sample, in this case, up to distances of 19 pc. The use of deeper surveys such as UKIDSS and VISTA will allow us to probe further distances and expand the number of such systems. Considering that 2MASS is complete up to a \textit{J} $=$ 16, and taking into account the number of L dwarfs found, one can expect that by going three orders of magnitude fainter in the aforementioned surveys we can expect to increase the volume of sky sampled by up to 30 times. This means we can expect to find around 500 times more L dwarfs having a MS star as companion in an all sky sample. Ultimately, it is by studying a large sample of UCDs with abundances and age constraints from their companions that we can further understand and improve current atmoshperic models of these objects. 

Finally we show that some of the L dwarfs present in systems with a MS star could be unresolved binaries, due to their relative overluminosity for their spectral type. One interesting case is the HD2057 system, in which two resolved L4 dwarfs might actually be unresolved binaries themselves, thus making this a possible quadruple L dwarf system with an F8 star as the primary.

\section{Acknowledgements}
J.Gomes and ACD-Jones are supported by RoPACS, a Marie Curie Initial Training Network funded by the European Commission's Seventh Framework Programme. ACD-J is also supported by a FONDECYT postdoctorado fellowship under project no. 3100098. DP, ZZ and HJ have received support from RoPACS during this research. This research has benefitted from the M, L, and T dwarf compendium housed at DwarfArchives.org and maintained by Chris Gelino, Davy Kirkpatrick, and Adam Burgasser. This publication makes use of data products from the 2MASS, which is a joint project of the University of Massachusetts and the IPAC/Caltech, funded by NASA and the National Science Foundation. We have also made use of data products from WISE, which is a joint project of the University of California, Los Angeles, and the Jet Propulsion Laboratory/California Institute of Technology, funded by the National Aeronautics and Space Administration. This research has made use of the SIMBAD data base, operated at CDS, Strasbourg, France. We thank Tom Marsh for the use of \textsc{Pamela} and \textsc{molly}. Based on observations made with the Italian Telescopio Nazionale Galileo (\textsc{tng}) operated on the island of La Palma by the Fundacin Galileo Galilei of the \textsc{inaf} (Istituto Nazionale di Astrofisica) at the Spanish Observatorio del Roque de los Muchachos of the Instituto de Astrofisica de Canarias.


\appendix

\bsp

\label{lastpage}

\end{document}